\documentclass[twocolumn,superscriptaddress,letter]{revtex4}%
\usepackage{amssymb}
\usepackage{color}
\usepackage{graphicx}
\usepackage{dcolumn}
\usepackage{bm}
\usepackage[header,title,page,titletoc]{appendix}
\usepackage{amsmath}
\usepackage{subfigure}
\usepackage{amsfonts}%
\providecommand{\U}[1]{\protect \rule{.1in}{.1in}}

\begin{document}
\title{Classification of Majorana Fermions in Two-Dimensional Topological Superconductors}
\author{Qiu-Bo Cheng}
\affiliation{Department of Physics, Beijing Normal University, Beijing, 100875, P. R. China}
\author{Jing He}
\affiliation{Department of Physics, Beijing Normal University, Beijing, 100875, P. R. China}
\affiliation{Department of Physics, Hebei Normal University, HeBei, 050024, P. R. China}
\author{Su-Peng Kou}
\thanks{Corresponding author}
\email{spkou@bnu.edu.cn}
\affiliation{Department of Physics, Beijing Normal University, Beijing, 100875, P. R. China}

\begin{abstract}
Recently, Majorana Fermions (MFs) have attracted intensive attention due to
their exotic statistics and possible applications in topological quantum
computation (TQC). They are proposed to exist in various two-dimensional (2D)
topological systems, such as $p_{x}+ip_{y}$ topological superconductor and
nanowire-superconducting hybridization system. In this paper, two types of
Majorana Fermions with different polygon sign rules are pointed out. A
\textquotedblleft smoking gun\textquotedblright \ numerical evidence to
identify MF's classification is presented through looking for the signature of
a first order topological quantum phase transition. By using it, several 2D
topological superconductors are studied.

\end{abstract}
\maketitle

\section{Introduction}

Majorana fermion is a real fermion that is its own
antiparticle\cite{Majorana,Wilczek,Martin}. Because of its exotic properties
and the possible exotic statistics\cite{Kitaev01,Fu08,Lutchyn10, Oreg10,
Sau10,Alicea10,Potter10,
Alicea11,Halperin12,Stanescu13,Mourik12,Das12,Deng12,Rokhinson12,Churchill13},
in condensed matter physics, the search for Majorana fermions (MFs) has
attracted increasing interests. A variety of schemes to realize MFs (more
accurately, Majorana bound states) have been proposed. A possible approach is
to create a quantized vortex ($\pi$-flux) in the $p_{x}+ip_{y}$-wave
topological superconductors (SC) that traps MFs in
vortex-core\cite{Moore91,Read00,Ivanov01,Nayak08,Alicea12}. Then, it is known
that the quantized vortex in two dimensional (2D) topological SC with nonzero
Chern number hosts a MF with exact zero energy. Another different approach to
realize MFs is to consider a one-dimensional (1D) electronic nano-structures
proximity-coupled to a bulk superconductor\cite{Kitaev01}, of which the
unpaired Majorana fermions appear as the end-states. Then, based on this idea,
several schemes are proposed to realize MFs that appear as the end-states of
line-defects in 2D non-topological SCs\cite{kou}.

To describe the Majorana zero mode, a real fermion field called Majorana
fermion $\gamma=\int d{r}[u_{0}\psi^{\ast}+v_{0}\psi]$ ($\gamma^{\dagger
}=\gamma$) is introduced. We consider a 2D gapped SC with a pair of Majorana
modes with nearly zero energies, of which the corresponding MFs are denoted by
$(\gamma_{1},$ $\gamma_{2})$. To describe the subspace of the system with two
nearly degenerate states, the Fermion-parity operator $\mathrm{\hat{P}%
}=-i\gamma_{1}\gamma_{2}$ is introduced. Since $\mathrm{\hat{P}}^{2}=1$,
$\mathrm{\hat{P}}$ has two eigenvalues $\pm1,$ called even and odd
Fermion-parities, respectively. Generally, there exists the coupling between
two MFs and the effective Hamiltonian is given by $it\gamma_{1}\gamma_{2}$
where $t$ is the coupling constant. The quantum systems with multi-MFs (we
call this lattice model to be \emph{Majorana lattice model}) show nontrivial
topological properties, including a nonvanishing Chern number, chiral Majorana
edge state\cite{VILL,kou2,kou1}. It was pointed out that the Majorana lattice
model is really an induced "topological superconductor" on the parent TSC.

A question arises, "\emph{Do MFs in different topological superconductors
belong to the same class?}" Without no detectable degree of freedoms (MFs have
zero energy, zero charge, zero spin), it is believed that MFs have "\emph{no
Hair}". Therefore, it was believed that all MFs in different models are same
and belong to the same class. In this paper we point out that in 2D
topological systems, MFs do have "hair", that is the number of quantized
vortex, a topological degree of freedom. As a result, there exist \emph{two
}universal classes of MFs: MFs binding a $\pi$-flux and those with no
flux-binding. We then introduce a topological value to characterize the two
classes of MFs and propose an (indirect) numerical approach to identify the
class of MFs in topological systems. Our basis is the fact that multi-MFs with
$\pi$-flux may be topologically different from those with no flux-binding in
the Fermion-parity of the ground states and the order of quantum phase
transitions. By calculating a dimensionless parameter, the gap ratio (see
discussion below), the quantum number of the MFs becomes an \emph{observable} "quantity".

\section{Classification of Majorana fermions}

We begin by giving the definitions of two classes of MFs in 2D topological
systems. One class contains the usual MFs without flux-binding (we call this
class \emph{normal MFs}); the other class contains composite objects of an MF
together with a $\pi$-flux (we call this class \emph{topological MFs}). See
the illustrations of the two classes of MFs in Fig.\ref{fig1}.(a) and
Fig.\ref{fig3}.(a).

We consider a system with coupled MFs. The coupling strength is just the
energy splitting from the intervortex quantum tunneling. We call this
effective description as the Majorana lattice model, of which the
tight-binding Hamiltonian can be written as
\begin{equation}
\mathcal{H}_{m.f}=i\sum_{(j,k)}s_{jk}t_{jk}\gamma_{k}\gamma_{j}%
\end{equation}
where $t_{jk}$ is the hopping amplitude from $j$ to $k$, and satisfies
$t_{jk}=t_{jk}^{\ast}$. $\gamma_{j}$ is a Majorana operator ($\gamma_{j}%
^{\dag}=\gamma_{j}$) obeying anti-commutate relation $\{ \gamma_{j},\gamma
_{k}\}=2\delta_{jk}$. $s_{ij}=-s_{ji}$ is a gauge factor. Thus, the total
number of Majorana modes $N$ must be even and then we can divide the Majorana
lattice into two sublattices. The pair $(j,k)$ denotes the summation that runs
over all the nearest neighbor (NN) pairs (with hopping amplitude $t$) and all
the next-nearest neighbor (NNN) pairs (with hopping amplitude $t^{\prime}$).
Each triangular plaquette possesses $\pm \pi/2$ quantum flux effectively.

This Hamiltonian allows different $Z_{2}$ gauge choices (sign rules)
$s_{jk}=\pm1$. From the theory of projective symmetry group, there are two
possible gauge choices (sign rules) that correspond to two different classes
of MFs: topological MFs obey a \emph{topological polygon sign rule;} normal
MFs obey a \emph{normal polygon sign rule}. According to topological polygon
sign rule for topological MFs, there exists an extra phase related to a closed
path that forms a polygon, given by half the sum of the interior angles of the
polygon\cite{EGR}. Thus, for $4N$ ($4N\pm1$) topological MFs on a ring, there
exists $\pm \pi$ ($\pm \pi/2$) flux inside the ring; for $4N\pm2$ topological
MFs on a ring, there is no flux inside the ring. On the contrary, according to
normal polygon sign rule for normal MFs, there also exists an extra phase
related to a closed path that forms a polygon. However, for $4N\pm1$ normal
MFs on a ring, there exists $\pm \pi/2$ flux inside the ring and for $4N$ or
$4N\pm2$ normal MFs on a ring, there is no flux inside the ring.

\section{Quantum phase transitions in dimerized Majorana rings}

\subsection{The Hamiltonian of Majorana rings}

We begin from a (dimerized) Majorana ring of $4N$ MFs ($N$ is a positive
integer number). The Hamiltonian reads\cite{Kitaev01}
\begin{align}
\mathcal{\hat{H}}_{\mathrm{MF}}  &  =i\sum_{j=1}^{2N-\delta_{1,N}}%
(J_{1}^{\prime}\gamma_{jb}\gamma_{j+1,a}+J_{1}\gamma_{ja}\gamma_{jb}%
\label{1}\\
&  +J_{2}\gamma_{jb}\gamma_{j+1,b}+J_{2}\gamma_{ja}\gamma_{j+1,a})\nonumber
\end{align}
where $a$, $b$ denote the sublattices in a unit cell, $J_{1}$ is the coupling
constants between two MFs in a unit cell, $J_{1}^{\prime}$ ($J_{2}$) are the
(next) nearest coupling constants between two MFs in different unit cells. For
the case of $N=1$, $\delta_{1,N}=1;$ For the case of $N>1$, $\delta_{1,N}=0$.
For usual systems, we have $J_{2}<J_{1},$ $J_{1}^{\prime}$.

For a Majorana ring with $4N$ normal MFs (we call it \emph{normal Majorana
ring}), due to normal polygon sign rule, we always have periodic boundary
condition; while for a Majorana ring with $4N$ topological MFs (we call it
\emph{topological Majorana ring}), due to topological polygon sign rule, we
have anti-periodic boundary condition owing to an extra $\pi$ flux inside the
ring. Fig.\ref{fig1}.(a) and Fig.\ref{fig3}.(a) illustrate a Majorana ring
with $8$ normal MFs and that with $8$ topological MFs, respectively.

\subsection{Quantum phase transitions in normal Majorana rings}

\begin{figure}[ptbh]
\includegraphics[width=0.5\textwidth]{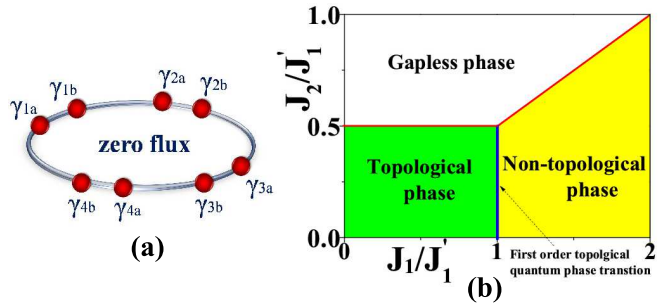}\caption{(color online) (a)
Illustration of an normal Majorana ring with 8 normal Majorana Fermions (red
dots); (b) Phase diagram of normal Majorana ring in thermodynamic limit,
$N\rightarrow \infty$. The blue line denotes the first order topological
quantum phase transition that switches Fermion-parity of the ground state and
the red lines denote the second order phase transitions.}%
\label{fig1}%
\end{figure}

\begin{figure}[ptbh]
\includegraphics[width=0.5\textwidth]{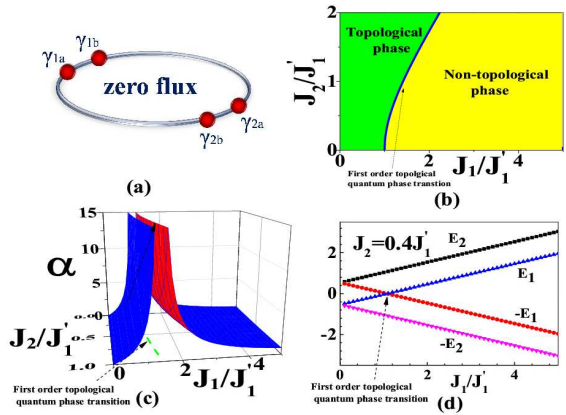}\caption{(color online) (a)
Illustration of the normal Majorana ring with four Majorana Fermions; (b)
Phase diagram of the normal Majorana ring with four Majorana Fermions. The
blue line denotes the first order topological quantum phase transition that
switches Fermion-parity of the ground state; (c) The gap-ratio of the normal
Majorana ring with four Majorana Fermions. The divergence of the gap-ratio,
$\alpha \rightarrow \infty,$ denotes a first order topological quantum phase
transition; (d) The energy levels of the normal Majorana ring with four
Majorana Fermions at $J_{2}=0.4J_{1}^{\prime}$. The level-crossing denotes a
first order topological quantum phase transition.}%
\label{fig2}%
\end{figure}

Firstly, we study a (dimerized) Majorana ring with $4N$ normal MFs. We pair
($\gamma_{ja}$, $\gamma_{jb}$) into a complex fermion as $\gamma_{ja}%
=(c_{j}+c_{j}^{\dag})/\sqrt{2},$ $\gamma_{jb}=-i(c_{j}-c_{j}^{\dag})/\sqrt
{2},$ where $c_{j}$ ($c_{j}^{\dag}$) annihilates (creates) a complex fermion.
Then the Majorana ring's energy spectra can be obtained through a fourier
transformation $c_{k}=N^{-1/2}\sum_{i}e^{-ikR_{i}}c_{i}$. In thermodynamic
limit, $N\rightarrow \infty$, the Hamiltonian in momentum space takes the form
of
\begin{align}
\mathcal{\hat{H}}_{\mathrm{MF}}  &  =\sum_{k}\psi_{k}^{\dag}[(-J_{1}^{\prime
}\cos k+J_{1})\tau^{z}\label{2}\\
&  +(J_{1}^{\prime}\tau^{y}-2J_{2})\sin k]\psi_{k}\nonumber
\end{align}
where $\psi_{k}^{\dag}=(c_{k}^{\dag},c_{-k})$ and $\tau^{z},$ $\tau^{y}$ are
Pauli matrices. The energy spectra are given by
\begin{equation}
E_{\pm}(k)=-2J_{2}\sin k\pm \sqrt{(J_{1}^{\prime})^{2}-2J_{1}^{\prime}J_{1}\cos
k+J_{1}^{2}}%
\end{equation}
where $k=\pi n/N,$ $n=1,2,...,2N.$

To characterize the topological properties of the ground states, we define a
Fermion-parity operator in the following form
\begin{equation}
\mathrm{\hat{P}}=\prod_{j=1}^{2N}(-i\gamma_{ja}\gamma_{jb}).
\end{equation}
Since $\mathrm{\hat{P}}^{2}=1$, $\mathrm{\hat{P}}$ has two eigenvalues $\pm1,$
called even and odd Fermion-parities, respectively. Due to $[\mathrm{\hat{P}%
},$ $\mathcal{\hat{H}}_{\mathrm{MF}}]=0$, the ground state $\left \vert
vac\right \rangle $ should have a determinant parity $\mathrm{\hat{P}%
}\left \vert vac\right \rangle =\pm \left \vert vac\right \rangle $. For the case
of $\mathrm{\hat{P}}\left \vert vac\right \rangle =-\left \vert vac\right \rangle
,$ the ground state is a topological phase; for the case of $\mathrm{\hat{P}%
}\left \vert vac\right \rangle =\left \vert vac\right \rangle ,$ the ground state
is a non-topological phase.

For the Majorana ring described by\ Eq.(\ref{1}), the eigenvalues of
$\mathrm{\hat{P}}$ is equal to
\begin{equation}
\mathrm{sgn}(E(k=0)\cdot E(k=\pi))
\end{equation}
$\ $where $E(k=0)=J_{1}-J_{1}^{^{\prime}}$ at $k=0$ and $E(k=\pi)=J_{1}%
+J_{1}^{^{\prime}}$ at $k=\pi.$ Thus, at $J_{1}=J_{1}^{^{\prime}}$, the energy
gap closes, $E_{\pm}(k)=0,$ at which a topological quantum phase transition
(TQPT) occurs\cite{Kitaev01}. In Fig.\ref{fig1}.(b), we plot the global phase
diagram of the normal Majorana ring in thermodynamic limit, $N\rightarrow
\infty$. There are three phases: gapless phase (we don't discuss this phase
due to triviality), topological phase and non-topological phase. The blue line
denotes the first order TQPT that switches the Fermion-parity of the ground
state\cite{Kitaev01} and the red lines denote the second order phase
transitions. In yellow region ($J_{1}>J_{1}^{^{\prime}}$, $J_{2}<J_{1}$), we
have $\mathrm{P}=1$ and the ground state corresponds to a trivial state
(non-topological phase) with even Fermion-parity, $\mathrm{\hat{P}}\left \vert
vac\right \rangle =\left \vert vac\right \rangle $. In green region ($J_{1}%
<J_{1}^{^{\prime}},$ $J_{2}<J_{1}^{\prime}$), we have $\mathrm{P}=-1$ and the
ground state corresponds to a topological phase with odd Fermion-parity,
$\mathrm{\hat{P}}\left \vert vac\right \rangle =-\left \vert vac\right \rangle
$\cite{kou3}. So, for a normal Majorana ring with \emph{arbitrary} $N,$ a
\emph{first} order TQPT that switches the Fermion-parity occurs at
$J_{1}=J_{1}^{^{\prime}}$.

On the other hand, for a Majorana ring with 4 normal MFs (or $N=1$), the phase
diagram (See Fig.\ref{fig2}.(b)) differs from the case in thermodynamic limit
(or $N\rightarrow \infty$) (See Fig.\ref{fig1}.(b)). The blue line in
Fig.\ref{fig2}.(b) denotes the first order TQPT that switches the
fermion-parity of the ground state. To make the TQPT in a Majorana ring with
only 4 MFs more clear, we plot the energy levels for the case $J_{2}%
=0.4J_{1}^{\prime}$ in Fig.\ref{fig2}.(d). Now, the four energy levels are
$E_{\pm}(k=0),$ $E_{\pm}(k=\pi).$ The level-crossing in Fig.\ref{fig2}.(d)
shows the first order TQPT corresponding to the blue line in Fig.\ref{fig2}.(b)).

In addition, to characterize the TQPT, we introduce a dimensionless parameter
- \emph{gap-ratio},
\begin{equation}
\alpha=\frac{\max \left \vert E_{k}\right \vert -\min \left \vert E_{k}\right \vert
}{\min \left \vert E_{k}\right \vert }%
\end{equation}
where $\max \left \vert E_{k}\right \vert $ and $\min \left \vert E_{k}\right \vert
$ are the maximum value and minimum value of the energy levels of the
multi-MFs, respectively. When the gap-ratio turns to infinite, the energy gap
closes and TQPT occurs. In the thermodynamic limit, the gap-ratio $\alpha$
becomes $\frac{W_{\mathrm{MF}}}{\Delta_{\mathrm{MF}}}$ where $W_{\mathrm{MF}%
}=\max \left \vert E_{k}\right \vert -\min \left \vert E_{k}\right \vert $ is the
band-width and $\Delta_{\mathrm{MF}}=\min \left \vert E_{k}\right \vert $ is the
energy gap of the Majorana ring. This is why we call $\alpha$ gap-ratio. On
the other hand, for an normal Majorana ring with four MFs (or $N=1$), due to
$\max \left \vert E_{k}\right \vert =\left \vert E(k=\pi)\right \vert $ and
$\min \left \vert E_{k}\right \vert =\left \vert E(k=0)\right \vert $, the
gap-ratio $\alpha$ is equal to $\left \vert \frac{E(k=\pi)}{E(k=0)}\right \vert
-1.$ In particular, at TQPT, accompanied by gap-closing, $\Delta_{\mathrm{MF}%
}=0$ or $E(k=0)=0$, the gap-ratio diverges,
\begin{equation}
\alpha \rightarrow \infty.
\end{equation}
See the results in Fig.\ref{fig2}.(c).

\subsection{Quantum phase transitions in topological Majorana rings}

\begin{figure}[ptbh]
\includegraphics[width=0.5\textwidth]{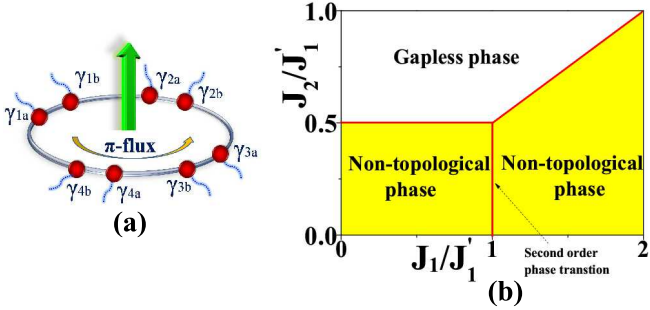}\caption{(color online)
Illustration of a topological Majorana ring with 8 topological Majorana
Fermions (red dots with strings). Due to the polygon rule, there exists an
extra $\pi$-flux inside the ring for topological Fermions; (b) Phase diagram
of topological Majorana ring in thermodynamic limit, $N\rightarrow \infty$. The
red lines denote the second order phase transitions.}%
\label{fig3}%
\end{figure}

\begin{figure}[ptbh]
\includegraphics[width=0.5\textwidth]{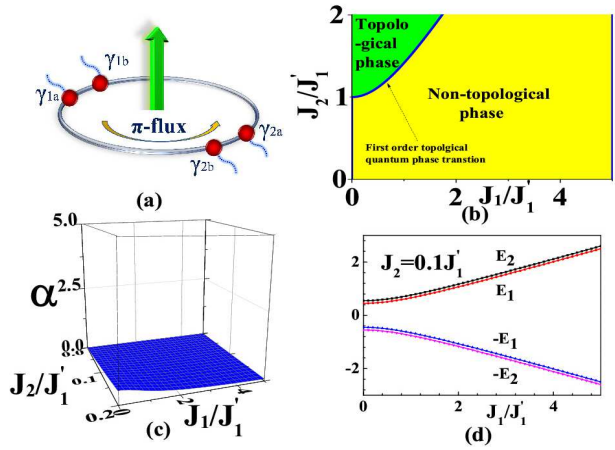}\caption{(color online) (a)
Illustration of the topological Majorana ring with four Majorana Fermions. Due
to the polygon rule, there exists an extra $\pi$-flux inside the ring for
topological Majorana Fermions; (b) Phase diagram of the topological Majorana
ring with four Majorana Fermions. The blue line denotes the first order
topological quantum phase transition that switches Fermion-parity of the
ground state; (c) The gap-ratio of the topological Majorana ring with four
Majorana Fermions; (d) The energy levels of the topological Majorana ring with
four Majorana Fermions at $J_{2}=0.1J_{1}^{\prime}$.}%
\label{fig4}%
\end{figure}

Next, we study a (dimerized) Majorana ring of $4N$ topological MFs. Due to the
extra $\pi$-flux inside the ring, we have an anti-periodic boundary condition
of the topological Majorana ring. The energy spectra are the same to those of
the normal Majorana ring as
\begin{equation}
E_{\pm}(k)=-2J_{2}\sin k\pm \sqrt{(J_{1}^{\prime})^{2}-2J_{1}^{\prime}J_{1}\cos
k+J_{1}^{2}}%
\end{equation}
with different wave-vectors, $k=\pi(n-1/2)/N,$ $n=1,2,...,2N.$\quad Now, there
are no high symmetry points at $k=0,$ $k=\pi$. As a result, the ground state
always has even fermion-parity, $\mathrm{\hat{P}}\left \vert vac\right \rangle
\equiv \left \vert vac\right \rangle $. In thermodynamic limit, a \emph{second}
order phase transition occurs at $J_{1}=J_{1}^{^{\prime}}$ between a
non-topological phase (the left yellow region in Fig.\ref{fig3}.(b)) and
another non-topological phase (the right yellow region in Fig.\ref{fig3}.(b)).
For the topological Majorana rings with finite $N$, the energy levels can
smoothly change from one phase to the other without level-crossing. At the
quantum critical point, $J_{1}=J_{1}^{^{\prime}}$, in the thermodynamic limit,
we have $\alpha \sim N\rightarrow \infty$; while for finite $N$, $\alpha$ is a
finite value.

On the other hand, for a Majorana ring with 4 topological MFs (or $N=1$), the
phase diagram \emph{differs} from that for the normal case. See the results in
Fig.\ref{fig4}.(b). A first order TQPT occurs in the region with large $J_{2}$
that is irrelevant to traditional topological systems. In the region with
small $J_{2}$, the first order TQPT will never occur. In Fig.\ref{fig4}.(c),
the four energy levels for a topological Majorana ring with four MFs are
$E_{\pm}(k=\pi/2),$ $E_{\pm}(k=-\pi/2).$ In Fig.\ref{fig4}.(d), the energy
levels of the topological Majorana ring with four Majorana Fermions are
plotted for the case of $J_{2}=0.1J_{1}^{\prime}$. The gap-ratio for
topological MFs is obtained to be
\[
\alpha=\left \vert \frac{E_{+}(k=-\pi/2)}{E_{+}(k=\pi/2)}\right \vert -1.
\]
For the case of $J_{2}\ll J_{1},$ $J_{1}^{\prime}$, due to $E_{\pm}%
(k=\pi/2)\simeq E_{\pm}(k=-\pi/2)$, we always have a very small gap-ratio as
\begin{equation}
\alpha \rightarrow0.
\end{equation}

\section{Numerical Method to identify the Majorana Fermion's classification}

From above discussion, the sharp distinctions between normal and topological
MFs are found in relevant physics ($J_{2}<J_{1},$ $J_{2}<J_{1}^{\prime}$): for
four normal MFs on a ring, at the point of first order TQPT, $\alpha
\rightarrow \infty$ (or $\alpha \gg1$); for four topological MFs on a ring,
without the first order TQPT, $\alpha \rightarrow0$ (or $\alpha<1$). Therefore,
we propose a numerical method to distinguish the normal/non-normal polygon
sign rule for the MFs by calculating the gap-ratio $\alpha$ in a topological
system with four MFs that form a (dimerized) Majorana ring.

\begin{figure}[ptbh]
\includegraphics[width=0.45\textwidth]{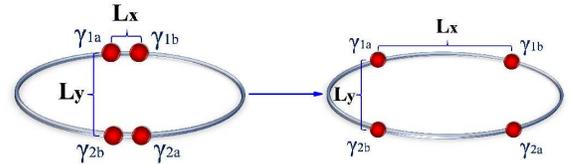}\caption{(color online)
Illustration of the varying of a Majorana ring with four Majorana Fermions
from the limit $L_{x}/L_{y}\ll1$ (or $J_{1}/J_{1}^{^{\prime}}\gg1$) to the
limit $L_{x}/L_{y}\gg1$ (or $J_{1}/J_{1}^{^{\prime}}\ll1$).}%
\label{fig5}%
\end{figure}

In the first step, we study the given 2D topological system without
considering the MFs. After diagonalizing the BdG equation, we obtain the
energy spectra in momentum space $E(\mathbf{k})$ and the energy gap of the
system $\Delta_{f}$.

In the second step, the energy levels of the 2D topological system with a pair
of MFs ($\gamma_{1}$, $\gamma_{2}$) are calculated by numerical approach. We
can derive the energy levels of the MFs with almost zero energies, $\pm E$,
($E>0$). When there are two MFs nearby, the quantum tunneling effect occurs
and the two MFs couple. The energy splitting between two nearly zero modes
$\Delta E=2E$ versus the distance $L$ of the two MFs can be obtained. In
general, $\Delta E$ oscillates and decreases exponentially with $L$ and can be
described by a function as $\Delta E\propto e^{-L/\xi}\left \vert \cos(L\cdot
k_{F})\right \vert $ where $\xi \sim v_{F}/\Delta_{f}$ is the correlated length
$\xi$ and $v_{F}$ is the Fermi velocity. We plot the enveloping line of
$\Delta E$ (we denote it by $\Delta E_{\mathrm{el}}$) by choosing the distance
to be $L_{\mathrm{el}}\sim \pi n/k_{F}$ where $n$ is an integer number. It is
obvious that $\Delta E_{\mathrm{el}}$ becomes a monotonous function via $L$
and decays exponentially, $\Delta E_{\mathrm{el}}\propto e^{-L/\xi}$.

In the third step, we consider the topological system with four localized MFs
that form an $L_{x}\times L_{y}$\ square (a dimerized Majorana ring). The
distance between $\gamma_{1a}$, $\gamma_{1b}$ and that between $\gamma_{2a}$,
$\gamma_{2b}$ is $L_{x}$, the distance between $\gamma_{1a}$, $\gamma_{2b}$
and that between $\gamma_{1b}$, $\gamma_{2a}$ is $L_{y}$. If we fix $L_{y}$
(or fix $J_{1}^{\prime}$), and then we can smoothly tune $J_{1}$ by changing
$L_{x}=L_{\mathrm{el}}\sim \pi n/k_{F}$. In the limit of $L_{x}\rightarrow
\infty$, we may have $J_{1}\ll J_{1}^{^{\prime}}$; in the limit of
$L_{x}\rightarrow0$, we may have $J_{1}\gg J_{1}^{^{\prime}}$. See the
illustration in Fig.\ref{fig5}. During varying $L_{x}/L_{y}$ (or $J_{1}%
/J_{1}^{^{\prime}}$ and eventually $\alpha$), we carefully look for the
signature of a first order TQPT with level-crossing that indicates a diverge
gap-ratio, or $\alpha \rightarrow \infty$. For a 2D topological system on
lattice, due to the discreteness of $L_{x}/L_{y}$ (or $J_{1}/J_{1}^{^{\prime}%
}$), the gap-ratio $\alpha$ will never diverge but may be a fairly large
value. The large gap-ratio (for example, $\max \alpha>10$) could be regarded as
a strong evidence of the first order TQPT. Eventually, the MFs obey normal
polygon sign rule. On the contrary, if the resulting gap-ratio $\alpha$ in a
2D topological system is always a small value (for example $\max \alpha<1$), we
can exclude the possibility of a first order TQPT and ultimately identify the
topological polygon sign rule of the MFs.

\section{Classify Majorana bound states in 2D topological superconductors}

In condensed matter systems, MFs are proposed to exist in various
two-dimensional (2D) topological SCs \cite{Kitaev01,Fu08,Lutchyn10, Oreg10,
Sau10,Alicea10,Potter10,
Alicea11,Halperin12,Stanescu13,Mourik12,Das12,Deng12,Rokhinson12,Churchill13}.
In 2D \emph{strong} topological SCs (so termed because of the non-zero Chern
number), MFs could be induced by quantized vortices\cite{Read00} or
dislocations\cite{dis,qi}; in 2D \emph{weak} topological SCs (so termed
because the Chern number is zero), MFs could be induced by line
defects\cite{Kitaev01,kou}. In the following, we studied MFs in strong
topological SCs in Sec.III.A and B and those in weak topological SCs in
Sec.III.C and D.

\subsection{Majorana bound states in a 2D $p_{x}+ip_{y}$ topological
superconductor}

\begin{figure}[ptbh]
\includegraphics[width=0.5\textwidth]{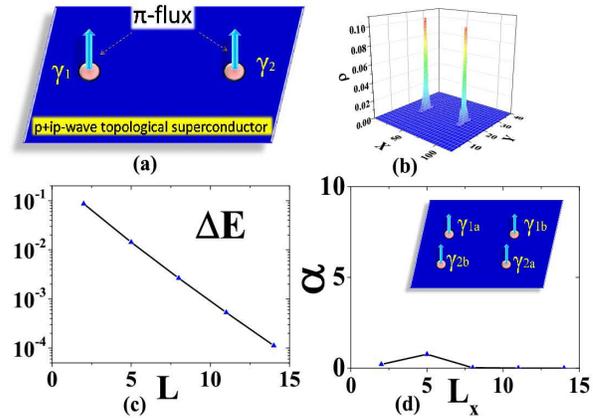}\caption{(color online) (a)
Illustration of two Majorana bound states trapped by $\pi$-fluxes in
$p_{x}+ip_{y}$ topological superconductor; (b) The particle density
distribution of Majorana zero modes around two $\pi$-fluxes; (c) The energy
splitting $\Delta E_{\mathrm{el}}$ via $L$ (the distance between the two
Majorana bound states); (d) The gap-ratio.}%
\label{fig6}%
\end{figure}

In the first example, we studied the MFs around vortices in a 2D $p_{x}%
+ip_{y}$ topological SC. The Hamiltonian of a $p_{x}+ip_{y}$ SC on a square
lattice is written as\cite{Read00}%
\begin{align}
\hat{H}_{\mathrm{pip}}  &  =-t\sum \limits_{j}\sum \limits_{\widehat{\mu
}=\widehat{x},\widehat{y}}(c_{j+\widehat{\mu}}^{\dagger}c_{j}+c_{j-\widehat
{\mu}}^{\dagger}c_{j})-\mu \sum \limits_{j}c_{j}^{\dagger}c_{j}\nonumber \\
&  +\sum \limits_{j}[\Delta(c_{j+\widehat{x}}^{\dagger}c_{j}^{\dagger
}+ic_{j+\widehat{y}}^{\dagger}c_{j}^{\dagger})+H.c.],
\end{align}
where $c_{j}$ is an electronic annihilation operator, $\mu$ is the chemical
potential, $\Delta$ is the SC pairing-order parameter and $t$ is the hopping
strength. The lattice constant was set to unity in this paper. In the
following, we chose the parameters as $t=1,$ $\mu=-1,$ $\Delta=0.4$. The
ground state was the weak-pairing phase that is a (strong) topological SC.

We first studied two MFs ($\gamma_{1}$, $\gamma_{2}$) around two vortices with
numerical calculations on a $120\times40$ lattice. We found that there exists
a Majorana zero mode around each vortex. Note the particle density around two
$\pi$-fluxes in Fig.\ref{fig6}.(b). When there are two fluxes nearby,
inter-flux quantum tunneling occurs and the two MFs couple. Fig.\ref{fig6}.(c)
shows the energy splitting $\Delta E_{\mathrm{el}}$. Next, we studied four
coupled MFs of vortices, $\gamma_{1a}$, $\gamma_{1b}$, $\gamma_{2a}$,
$\gamma_{2b},$ that formed an $L_{x}\times L_{y}$\ square (a dimerized
Majorana ring). See inset in Fig.\ref{fig6}.(d). By fixing $L_{y}$\ at $6$ and
varying $L_{x}$, we calculated the gap ratio $\alpha$ and show the result in
Fig.\ref{fig6}.(d), in which one can see a very small gap ratio. There is no
evidence of the first order TQPT. Therefore, we identified the MFs induced by
the vortices in $p_{x}+ip_{y}$ topological SCs to be \emph{topological} MFs.
It is obvious that this conclusion (topological MFs in 2D $p_{x}+ip_{y}$
topological SCs) is consistent with earlier results\cite{kou1}.

\subsection{Majorana bound states in an $s$-wave topological superconductor
with Rashba spin-orbital coupling}

\begin{figure}[ptbh]
\includegraphics[width=0.5\textwidth]{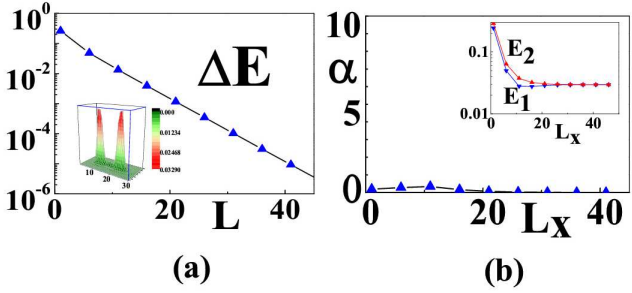}\caption{(color online) (a) The
energy splitting $\Delta E_{\mathrm{el}}$ via the length $L$ of two vortices.
The inset in (a) shows the particle density distribution of the Majorana bound
states; (b) The gap-ratio of four Majorana bound states. The inset in (b)
shows two positive energy levels of four coupled Majorana bound states.}%
\label{fig10}%
\end{figure}

The second model is MFs in an $s$-wave SC with Rashba spin-orbital (SO)
coupling on a square lattice\cite{MS,Lutchyn10,Oreg10,Sau10}. The Hamiltonian
is given by $\hat{H}_{\mathrm{s-wave-SO}}=\hat{H}_{\mathrm{kin}}+\hat
{H}_{\mathrm{so}}+\hat{H}_{\mathrm{sc}}$ where the kinetic energy term
$\hat{H}_{\mathrm{kin}}$, the Rashba SO coupling term $\hat{H}_{\mathrm{so}}$,
and the SC pairing term $\hat{H}_{\mathrm{sc}}$ are given as
\begin{align}
\hat{H}_{\mathrm{kin}}  &  =-t%
{\displaystyle \sum \limits_{j\sigma}}
{\displaystyle \sum \limits_{\widehat{\mu}=\widehat{x},\widehat{y}}}
(c_{j+\widehat{\mu}\sigma}^{\dagger}c_{j\sigma}+c_{j-\widehat{\mu}\sigma
}^{\dagger}c_{j\sigma})\\
&  -\mu%
{\displaystyle \sum \limits_{j\sigma}}
c_{j\sigma}^{\dagger}c_{j\sigma}-h%
{\displaystyle \sum \limits_{j}}
(c_{j\uparrow}^{\dagger}c_{j\uparrow}-c_{j\downarrow}^{\dagger}c_{j\downarrow
}),\nonumber \\
\hat{H}_{\mathrm{so}}  &  =-\lambda%
{\displaystyle \sum \limits_{j}}
[(c_{j-\widehat{x}\downarrow}^{\dagger}c_{j\uparrow}-c_{j+\widehat
{x}\downarrow}^{\dagger}c_{j\uparrow})\nonumber \\
&  +i(c_{j-\widehat{y}\downarrow}^{\dagger}c_{j\uparrow}-c_{j+\widehat
{y}\downarrow}^{\dagger}c_{j\uparrow})]+H.c.,\nonumber \\
\hat{H}_{\mathrm{sc}}  &  =-\Delta%
{\displaystyle \sum \limits_{j}}
(c_{j\uparrow}^{\dagger}c_{j\downarrow}^{\dagger}+H.c.).\nonumber
\end{align}
Here, $c_{j\sigma}$ ($c_{j\sigma}^{\dag}$) annihilates (creates) a fermion at
site $j=(j_{x},$ $j_{y})$ with spin $\sigma=(\uparrow,$ $\downarrow)$,
$\widehat{\mu}=\widehat{x}$ or $\widehat{y}$, which is a basic vector for the
square lattice. $\lambda$ serves as the SO coupling constant and $\Delta$ as
the s-wave SC pairing-order parameter. $\mu$ is the chemical potential and $h$
is the strength of the Zeeman field. The lattice constant was set as unity.
The parameters were chosen to be\ $t=1,$ $\mu=-4,$ $h=0.8,$ $\lambda=0.5,$
$\Delta=0.5.$ In this case, the ground state was a (strong) topological SC.

We then studied two MFs around two vortices through numerical calculations on
a $100\times24$ lattice. The particle distribution of the Majorana zero mode
is given in the inset in Fig.\ref{fig10}.(a). The results of the energy
splitting $\Delta E_{\mathrm{el}}$ via the distance of the two vortices $L$
are given in Fig.\ref{fig10}.(a). Next we studied four coupled MFs around the
vortices $\gamma_{1a}$, $\gamma_{1b}$, $\gamma_{2a}$, and $\gamma_{2b}$, that
formed an $L_{x}\times L_{y}$\ square (a dimerized Majorana ring). We fixed
$L_{y}$\ to be $6$ and varied $L_{x}$. The gap ratio is shown in
Fig.\ref{fig10}.(b). One can see a small gap ratio. As a result, we conclude
that MFs in an $s$-wave topological superconductor with Rashba SO coupling
obey topological polygon sign rule\cite{kou2}.

\subsection{Majorana bound states in a nanowire-SC hybridization system}

\begin{figure}[ptbh]
\includegraphics[width=0.5\textwidth]{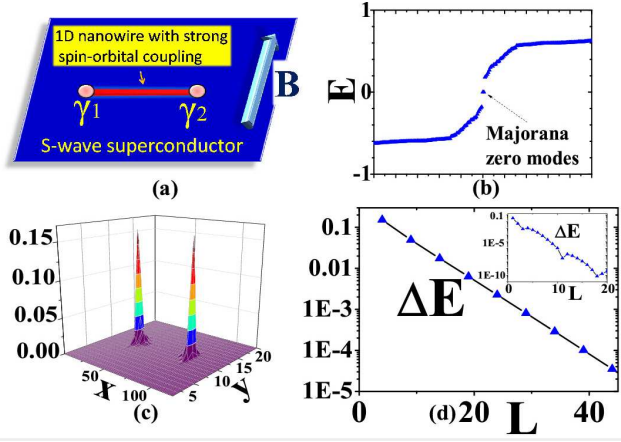}\caption{(color online) (a)
Illustration of nanowire-SC hybridization system; (b) Two Majorana bound
states with zero energy; (c) Particle distribution of Majorana bound states at
the ends of the nanowire on s-wave SC; (d) The energy splitting of two
Majorana bound states. The inset in (d) shows a very tiny $J_{2}$.}%
\label{fig7}%
\end{figure}

The third model was a 1D semiconducting nanowire with strong spin-orbital
coupling in a Zeeman field, proximity-coupled to an s-wave superconductor. See
the illustration in Fig.\ref{fig7}.(a). The Hamiltonian of the system is
$\hat{H}_{\mathrm{nano-SC}}=\hat{H}_{\mathrm{1D}}+\hat{H}_{\mathrm{sc}}$,
where $\hat{H}_{\mathrm{1D}}$ describes a 1D semiconducting nanowire; it is
written as
\begin{align}
\hat{H}_{\mathrm{1D}}  &  =-t_{s}\sum \limits_{j\sigma}\sum \limits_{\widehat
{\mu}=\widehat{x}}(c_{j+\widehat{\mu}\sigma}^{\dagger}c_{j\sigma
}+c_{j-\widehat{\mu}\sigma}^{\dagger}c_{j\sigma})+H.c.\\
&  -\lambda \sum \limits_{j}(c_{j-\widehat{x}\downarrow}^{\dagger}c_{j\uparrow
}-c_{j+\widehat{x}\downarrow}^{\dagger}c_{j\uparrow})-h\sum \limits_{j}%
(c_{j\uparrow}^{\dagger}c_{j\uparrow}-c_{j\downarrow}^{\dagger}c_{j\downarrow
})\nonumber \\
&  -\Delta \sum \limits_{j}c_{j\uparrow}^{\dagger}c_{j\downarrow}^{\dagger}%
-\mu \sum \limits_{j\sigma}c_{j\sigma}^{\dagger}c_{j\sigma},\nonumber
\end{align}
$\hat{H}_{\mathrm{sc}}$ describes the 2D superconductor out of 1D
semiconducting nanowire and is written as%
\begin{align}
\hat{H}_{\mathrm{sc}}  &  =-t\sum \limits_{j\sigma}\sum \limits_{\widehat{\mu
}=\widehat{x},\widehat{y}}(c_{j+\widehat{\mu}\sigma}^{\dagger}c_{j\sigma
}+c_{j-\widehat{\mu}\sigma}^{\dagger}c_{j\sigma})\nonumber \\
&  -\mu \sum \limits_{j\sigma}c_{j\sigma}^{\dagger}c_{j\sigma}-\Delta
\sum \limits_{j}(c_{j\uparrow}^{\dagger}c_{j\downarrow}^{\dagger}+H.c.),
\end{align}
where $c_{j\sigma}$ is an electronic annihilation operator and $t$ $(t_{s}),$
$\mu,$ $\Delta,$ $\lambda,$ $h$ denote the hopping parameters, the chemical
potential, the (induced) pairing order parameters, the spin-orbit coupling
strength and the Zeeman field, respectively. The lattice constant was set to
be unity. In the following, we chose the parameters as $t_{s}=1,$ $h=0.8,$
$\mu=-2,$ $\lambda=0.5,$ $\Delta=0.4,$ $t=0.4.$ The ground state of the system
is a (weak) topological SC.

\begin{figure}[ptbh]
\includegraphics[width=0.5\textwidth]{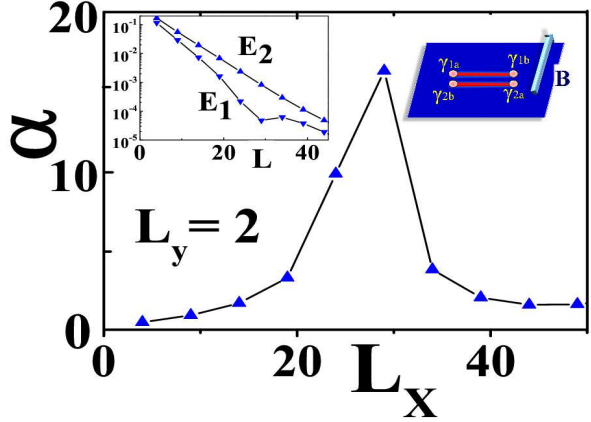}\caption{(color online) The
gap-ratio of the four Majorana bound states in nanowire-SC hybridization
system. The right inset illustrates two semiconducting nanowires on an SC
system. The left inset shows two positive energy levels of four Majorana bound
states. }%
\label{fig8}%
\end{figure}

Then, treating a nanowire by numerical calculations on a $120\times20$
lattice, we found that there are two Majorana zero modes ($\gamma_{1}$,
$\gamma_{2}$) at two ends of a nanowire. See the illustration in
Fig.\ref{fig7}.(a). There exist two zero modes in the energy gap shown in
Fig.\ref{fig7}.(b). In Fig.\ref{fig7}.(c), we plot the particle distribution
of the zero modes. The relationship between the energy splitting $\Delta
E_{\mathrm{el}}$ and the length $L$ of the nanowire is shown in Fig.\ref{fig7}%
.(d). Next, we considered two parallel 1D semiconducting nanowires on an SC.
See the right inset in Fig.\ref{fig8}. Here the low energy physics is
dominated by four coupled MFs, $\gamma_{1a}$, $\gamma_{1b}$, $\gamma_{2a}$,
$\gamma_{2b},$ that form an $L_{x}\times L_{y}$\ square (or a dimerized
Majorana ring). We fixed the distance between two parallel nanowires to be
$L_{y}=2$ and then changed the length $L_{x}$ of the two nanowires. The left
inset of Fig.\ref{fig8} shows the two positive energy levels from four MFs of
the two nanowires vs. $L_{x}$. In Fig.\ref{fig8}, the gap ratio $\alpha$ is
obtained. From Fig.\ref{fig8}, one can see that the maximum value of $\alpha$
reached $17$ at $L_{x}=30$. The sharp enhancement of the gap ratio $\alpha$ is
obviously the consequence of a first order TQPT. As a result, we conclude that
the MFs in the nanowire-SC hybridization system obey normal polygon sign rule.

\subsection{Majorana bound states in a p-wave superconductor on a honeycomb
lattice}

\begin{figure}[ptbh]
\includegraphics[width=0.5\textwidth]{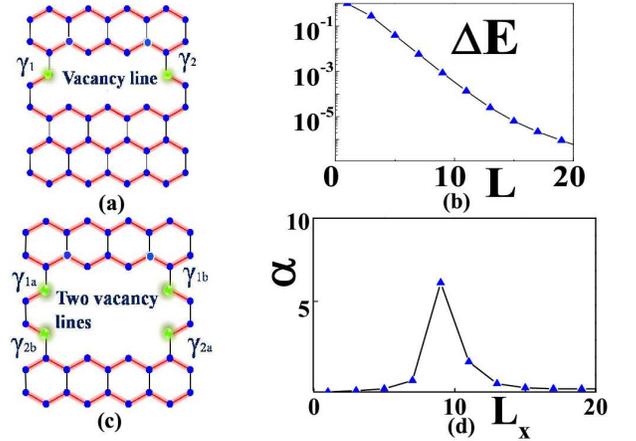}\caption{(color online) (a)
Illustration of two Majorana bound states induced by a line defect; (b) The
energy splitting $\Delta E_{\mathrm{el}}$ via the length $L$ of the line
defect; (c) Illustration of four Majorana bound states induced by two line
defects; (d) The gap-ratio of the four coupled Majorana bound states induced
by two parallel line defects.}%
\label{fig9}%
\end{figure}

The fourth model was MFs in a 2D p-wave superconductor on a honeycomb lattice.
The Hamiltonian of a p-wave superconductor for spinless fermions on a
honeycomb lattice is written as\cite{kou}
\begin{equation}
\hat{H}_{\mathrm{p-wave}}=-t\sum \limits_{\left \langle ij\right \rangle }%
c_{i}^{\dagger}c_{j}-t^{\prime}\sum \limits_{\left \langle \left \langle
ij\right \rangle \right \rangle }c_{i}^{\dagger}c_{j}-\sum \limits_{\left \langle
ij\right \rangle }\Delta_{ij}c_{i}^{\dag}c_{j}^{\dagger}+H.c.,
\end{equation}
where $t$ $(t^{\prime})$ denote the strengths of nearest (next nearest)
neighbor hopping. The p-wave pairing order parameters are defined by
$\Delta_{j,j+\mathbf{a}_{1}}=-\Delta_{j,j+\mathbf{a}_{2}}=\Delta$,
$\Delta_{j,j+\mathbf{a}_{3}}=0$. $\mathbf{a}_{\alpha}$ $(\alpha=1,2,3)$
denotes a vector that connects the nearest neighbor sites $i$ and
$i+\mathbf{a}_{\alpha}.$ Along the red links in Fig.\ref{fig9}.(a) and
Fig.\ref{fig9}.(c), the SC order parameter is finite; along black links, the
SC order parameter is zero. The lattice constant is set to be unity. We chose
$t=1,$ $t^{\prime}=0.01,$ $\Delta=1.34$ in this section. Here the ground state
is a (weak) topological SC.

In Ref.\cite{kou}, it was found that there exist two Majorana modes localized
at the ends of the line defect (LD), $\gamma_{1}$, $\gamma_{2}$. See the
illustration in Fig.\ref{fig9}.(a). The end of an LD can be considered to be
the boundary of a one-dimensional p-wave SC\cite{Kitaev01}. Thus, each end of
the LD traps a dangling Majorana zero mode. We studied two MFs around a LD
with numerical calculations on a $100\times50$ lattice and give the results of
the energy splitting $\Delta E_{\mathrm{el}}$ via the length $L$ of the LD in
Fig.\ref{fig9}.(b). Then, as shown in Fig.\ref{fig9}.(c), we studied four
coupled MFs of two parallel LDs that formed an $L_{x}\times L_{y}$\ square (a
dimerized Majorana ring). The distance between the two parallel LDs was fixed
to be $2$ (or $L_{y}=2$). We varied the length of the LDs, $L_{x}$. The
results of the gap ratio $\alpha$ are given in Fig.\ref{fig9}.(d), in which
the maximum value of $\alpha$ reaches $6.5$ at $L_{x}=9$. These results
indicate a first order TQPT with level-crossing. Thus, the MFs in p-wave
superconductors on honeycomb lattice also obey normal polygon sign rule.

\section{Discussion and conclusion}

\begin{figure}[ptbh]
\includegraphics[width=0.5\textwidth]{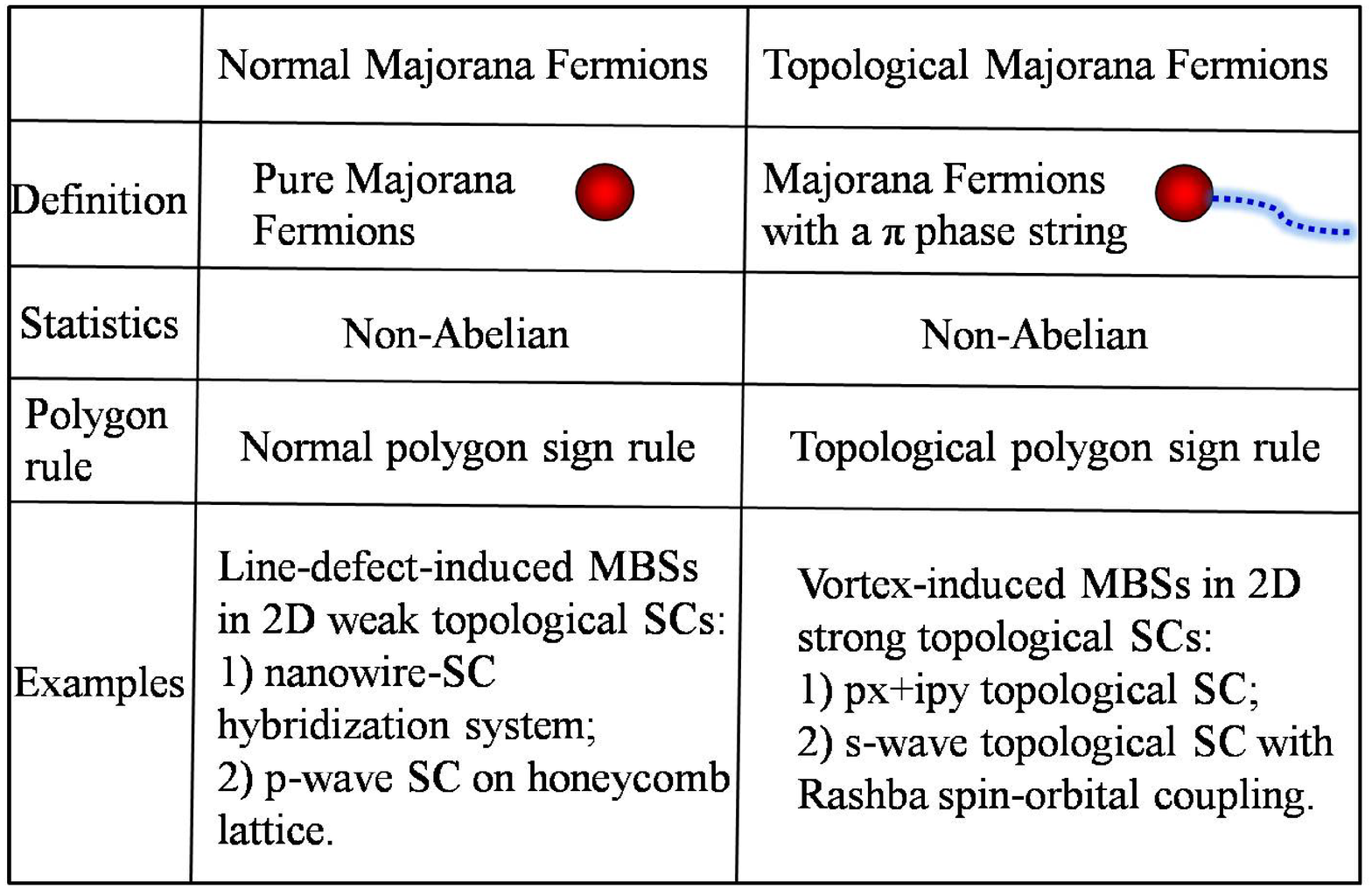}\caption{(color online) The
difference between normal Majorana Fermions and topological Majorana
Fermions.}%
\label{table}%
\end{figure}

In the end, we draw our conclusions (see the summary in Fig.11). We have
pointed out that in 2D topological superconductors, there exist two classes of
MFs: MFs obeying normal polygon sign rule and MFs obeying topological polygon
sign rule. A numerical approach was proposed to identify the polygon sign rule
of the MFs by looking for the signature of a first order TQPT of multi-MFs.
Applying the approach to study several 2D topological systems, we found that
vortex-induced MFs in 2D strong topological SCs (a $p_{x}+ip_{y}$ topological
superconductor in Sec.VI.A and an s-wave topological superconductor with
Rashba spin-orbital coupling in Sec.VI.B) obey topological polygon sign rule
and line-defect-induced MFs in 2D weak topological SCs (nanowire-SC
hybridization system in Sec.VI.C, p-wave superconductor on honeycomb lattice
in Sec.VI.D) obey normal polygon sign rule.

\begin{center}
{\textbf{* * *}}
\end{center}

This work is supported by National Basic Research Program of China (973
Program) under the grant No. 2011CB921803, 2012CB921704 and NSFC Grant No.
11174035, 11474025 and SRFDP.

\end{document}